\documentstyle[12pt]{article} \topmargin=-0.4in

\begin{document} 
\begin{flushright}
UCTP-120-99
\end{flushright}
\vskip0.2in
\begin{center} {\Huge \bf  Vortex solutions in nonabelian Higgs
theories}\\ \vskip0.1in 
Peter Suranyi\footnote{e-mail: suranyi@physics.uc.edu}\\
University of Cincinnati, Cincinnati, Ohio 45221, USA
 \end{center}

\abstract{ A new class of vortex solutions is found in $SU(2)$ gauge theories
with two adjoint representation Higgs bosons. 
Implications of these new solutions and their possible connection with Center
Gauge fixed pure gauge theories are discussed.}

The classification of vortex solutions in $SU(N)$ gauge theories requires
the investigation of mapping rotations around the vortex
axis ($U(1)$ group) into $SU(N)$. Unfortunately, all such maps are
continuously deformable into the trivial map. If one considers the group
$SU(N)/Z_N$ instead of $SU(N)$ then one finds nontrivial homotopy classes
classified by the abelian group, $Z_N$.  Then vortex solutions of
finite free energy (but infinite energy)  could exist  in four-dimensional
 and   finite energy
solitons  could exist in three-dimensional $SU(N)/Z_N$ gauge
theories.~\cite{thooft1} A prime candidate for finding such vortex solutions
would be an $SU(N)$ gauge theory with adjoint representation Higgs bosons. 

 Vortex solutions in nonabelian Higgs theories were
found some time ago by de Vega and Schaposnik.~\cite{devega1}~\cite{devega2}
~\cite{lozano1}~\cite{lozano2} In view of recent interest in vortex solutions
in center gauge fixed nonabelian gauge
theories~\cite{deldebbio}\cite{deldebbio2}\cite{alexandrou} the existence and
properties of vortex solutions is of considerable interest. After the discussion
of the new class of vortex solutions we will
point out a possible relationship between the two kinds of vortex solutions.

Higgs theories are defined by the Higgs potential.  The Higgs potential is not
unique for theories with multiple Higgs bosons. The existence of vortex
solutions requires the interaction the Higgs bosons with each other.  These
interactions are designed to keep the Higgs bosons non-parallel at infinite
distance away from the vortex line so that they would be able to break the gauge
symmetry  completely.  In previous work a simple mutual interaction, forcing the
Higgses to be orthogonal at infinity was used \begin{equation}
V(\Phi)=\frac{g}{16} [{\rm Tr}(\Phi^1\Phi^2)]^2+\dots, \label{twohiggs}
\end{equation}
 where we omit terms depending on a single Higgs field only. A
Higgs theory with such an interaction potential allows  for classical solutions
with orthogonal Higgs fields.  There are vortex
solutions in gauge fixed pure lattice gauge theories~\cite{alexandrou}
that, as we will point out later,   can be related to a gauge theory
with pair of adjoint Higgs fields that are {\em not orthogonal to each other}. 
Therefore, it is of considerable interest to investigate the existence of
vortex solutions in a wider class of Higgs theories that allow non-orthogonal
Higgs bosons.  In particular we will generalize (\ref{twohiggs}) such that the
interaction term has the form
 \begin{equation} V(\Phi)=\frac{g}{4} \left[\frac{1}{2}{\rm
Tr}(\Phi^1\Phi^2)-c\right]^2+\dots \label{twohiggs} \end{equation}
where $c$ is the cosine of the ``angle'' between the two Higgs bosons at
infinite distance away from the vortex line.  It turns out that such a
generalization requires the generalization of the
ansatz~\cite{devega1}~\cite{devega2} for the Higgs boson solutions. 

Though
much of setting up the problem parallels  Ref.~\cite{devega1}, for completeness,
and for establishing notation, we will provide a more or less complete
derivation. Though at first we will consider  an
$SU(N)$ gauge theory, for simplicity, calculations will be restricted to $N=2$.
We hope to return to the case of $N>2$ in a future publication.

 Due to the nontrivial fundamental group of
$SU(N)/Z_N$   
  classical solutions corresponding to gauge
transformations that vary smoothly
from one element of the center, $Z_N$, to another one, as one goes 
around a vortex line on a large circle, are stable. Since the elements of the
center are smoothly connected through elements of the Cartan subgroup,
$U(1)^{N-1}$, vortex configurations can only appear if the $SU(N)/Z_N$ part of
the gauge is fixed completely.  

 An adjoint Higgs boson is represented by a self adjoint $N\times N$
$SU(N)$ matrix that can always be diagonalized. Gauge transformations that
commute with this diagonal matrix form a $U(1)^{N-1}$ Cartan subgroup of the
gauge group. In other words, one Higgs boson in the adjoint representation fixes
only the gauge group to its Cartan subgroup.  Therefore, at least two,
non-parallel, adjoint Higgs bosons are required to fix
$SU(N)/Z_N$ completely. Accordingly, we set out to search for vortex solutions
in a Higgs theory with two adjoint Higgs bosons.  To simplify the algebra we
will restrict this part of the discussion to $SU(2)$. 

 As we study
time and z coordinate independent solutions our task is to minimize
the Hamiltonian,
 \begin{equation}
H=\int d^2x\, \mbox{\LARGE\{}\frac{1}{4}\vec
G_{\mu\nu}\vec G_{\mu\nu}+\frac{\Phi_0^2}{2}\sum_{s=1}^2
[D_\mu\vec\Phi^{(s)}]^2+ 
\frac{\lambda_1\Phi_0^4}{8} 
\sum_{s=1}^2\left(\vec\Phi^{(s)2}-1\right)^2+\frac{\lambda_2\Phi_0^4}{4}
\left(\vec\Phi^{(1)}\vec\Phi^{(2)}-c\right)^2\mbox{\LARGE\}},\label{lagrangian}
\end{equation}
to find vortex solutions.  In (\ref{lagrangian}) 
 $c$ is the "angle" between asymptotic fields, and as such, it should be
chosen to be in the interval $-1\le c\le1$.
$\Phi_0$ represents the vacuum expectation value of the Higgs fields.
The rescaled, adjoint representation Higgs boson fields, $\vec\Phi^{(s)}$, are
written in a three-vector form. 
 
The coupling $\lambda_2$ is needed to break the $U(1)$ (Cartan) subgroup of
$SU(2)$.  Notice that at $\lambda_2=0$ or at $\lambda_2\not=0$ and $c=\pm1$ 
this symmetry is not broken as the  solution of minimal energy is  obtained when
$\vec\Phi^{(1)}=\pm\vec\Phi^{(2)}$ and then gauge rotations around
their common direction are symmetries of the system. 

 By allowing $c\not=0$ (\ref{lagrangian}) is a generalization of the  
Hamiltonian of Ref.~\cite{devega1}~\cite{devega2}  As we will see later, this
generalization leads to  Higgs fields that rotate as a function of $r$.  At the
same time, to simplify the algebra and concentrate only on the effect of
the new constant $c$, we set the self-coupling and the vacuum
expectation value of the two Higgs bosons equal. 

 The gauge fields are related to the vector potential as 
\begin{equation} 
\vec G_{\mu\nu}=\partial_\mu \vec W_\nu-\partial_\nu\vec
W_\mu +e\vec W_\mu\times \vec W_\nu. \label{gauge} \end{equation}
The covariant derivative of the Higgs fields is defined as
\begin{equation}
D_{\mu}\vec\Phi^{(s)}=\partial_\mu \vec\Phi^{(s)}+e\vec W_\mu\times
\vec\Phi^{(s)}. \label{covariant}
\end{equation}

 We
fix the gauge by setting $\vec W_0=\vec W_3=0$ and  by imposing the gauge
condition \begin{equation} \partial_\alpha \vec W_\alpha=0,
\label{gaugefix}
\end{equation}
where the subscript $\alpha$ runs over $\alpha=1,2$.
The general form of the vector potential, satisfying (\ref{gaugefix}) that 
is finite at the origin is
\begin{equation}
\vec W_\alpha= \epsilon_{\alpha\beta}\,x_\beta \,\vec w(r) .
\label{vectorpot}
\end{equation}
We will discuss boundary conditions at $r=\infty$ later.
 
Note
that though the topological background is different, the form of the vector
potential is very similar to the vortex solution in  abelian
Higgs gauge theory.~\cite{nielsen}~\cite{mandelstam} The vortices of $U(1)$
gauge theory are nothing else but the covariant forms of magnetic flux tubes in a
superconductor.  Thus, the model we consider represents a non-abelian
superconductor.

We will use a singular gauge transformation, $U(\phi)$, in the ansatz for the
Higgs bosons. The fundamental
group of $SU(2)/Z_2$ is $Z_2$ so there is only one nontrivial homotopy class that
corresponds to gauge orbits that connect two opposite points in the $S_4$
representation of $SU(2)$. 
We choose $U(\phi)$ such that it belongs to the nontrivial homotopy
class. In other words, we will seek solutions of the field equations of the form
\begin{equation} \vec\Phi^{(s)}\cdot\vec\sigma=U(\phi)\, \vec\psi^{(s)}(r)
\cdot\vec\sigma\, U^\dagger(\phi),
\label{higgsansatz}
\end{equation}
where $\phi$ is the angle and $r$ is the radial distance in the $xy$ plane. The
singular gauge transformation, $U(\phi)$, is defined as
\begin{equation}
U(\phi)=e^{i\,\phi\, \vec\sigma\cdot \hat a/2},
\label{gauge-tr}
\end{equation}
where $\hat a$ is a constant unit vector.  When
$\phi$ runs from 0 to
$2\pi$ the gauge transformation $U(\phi)$ runs from the identity, $I$, to the
nontrivial element of the center, $-I$. Using gauge transformation we can
always choose the unit vector $\hat a$ to point in the direction of the
positive 3rd axis. 

Clearly, the gauge transformation, $U(\phi)$, cannot be deformed continuously
to identity.  Have we chosen $U(\phi)$ with twice the phase ($\phi\to2\phi$),
this would not be true. Then the transformation would have gone around a large
circle of the $S_4$ sphere when $\phi$ varies from 0 to $2\pi$, and such a circle
could be shrunk to a point through a series of continuous deformations.  

The boundary conditions at $r=0$
require that
$\vec\psi^{(s)}(0)=\lambda \hat a$, otherwise $\Phi^{(s)}$ would be singular at
the origin and $\int dr \,r (\partial_\mu \Phi)^2 $ would diverge. The finiteness
of the self-interaction terms of  (\ref{lagrangian}) also requires that 
$[\vec\psi^{(s)}(\infty)]^2=1$ and
$\vec\psi^{(1)}(\infty)\cdot\vec\psi^{(2)}(\infty)=c$.

Note that  (\ref{vectorpot}) implies that $\vec W_\mu\times
\vec W_\nu=0$. Then the field equations simplify considerably. The field
equation for the gauge field after multiplying with $U(\phi)$ from the right
and $U^\dagger(\phi)$ from the left will contain terms such as $U^\dagger
\,\vec w\cdot\vec \sigma\, U$ and $\vec w\cdot\vec \sigma$, multiplied by
functions dependent on $r$ only.  These terms have the same $\phi$
dependence only if $\vec w\cdot\vec \sigma, U]=0$.  In other words, $\vec w$ must
be parallel to $\hat a$. We will define the function $w$ by the relation $\vec
w=w\hat a.$ 

Before we proceed with minimizing the Hamiltonian we must make sure that 
our ansatz is consistent with the field equations.  We have seen already that
the $\phi$ dependence of the field equations cancels with the choice of $\vec w
= w \hat a.$ All three of the field equations are vector equations in
isotopic space.  As the gauge fields themselves are of the form $\vec w=w\hat
a$, the equations are consistent with structure if every term is of this
form.  Suppressing the argument $r$ we arrive at the condition 
\begin{equation}
\sum_s\vec{\psi^{(s)}}'\times\vec\psi^{(s)}+ew\sum_s\vec\psi^{(s)}
\times[\vec\psi^{(s)}\times \hat a] \sim \hat a. \label{constraint1}
\end{equation} The components of the terms on the left hand side of
(\ref{constraint1}), orthogonal to $\hat a$ are bilinear in the components of the
two Higgs field,  but linear in their components parallel to $\hat a$ and also
linear in their orthogonal components. Non-trivial solutions of
(\ref{constraint1}) are offered by the choices 
\begin{equation}\vec \psi^{(s)}_{a}=\pm
\vec \psi^{(2)}_{ a},\,\,\, \vec \psi_\bot^{(1)}=\mp \vec\psi_\bot^{(2)},
\label{soln}
\end{equation} 
where $\vec\psi_\bot^{(s)}$ and $\vec\psi_{a}^{(s)}$ denotes the components
of the Higgs fields orthogonal and parallel to $\hat a$, respectively. 

The field equations for
$\vec\psi^{(1)}(r)$ and $\vec\psi^{(2)}(r)$ have the form
  \begin{equation}
{\vec\psi^{(1)}}"+\vec
\frac{1}{r}\vec\psi^{(1)}{}'-e^2r^2\left(w+\frac{1}{e\,r^2}\right)^2\vec
\psi_\bot^{(1)}-\frac{\lambda_1}{2}\Phi_0^2\,\vec\psi^{(1)}[\vec\psi^{(1)2}-1] -
\frac{\lambda_2}{2}\Phi_0^2\,\vec\psi^{(2)}[\vec\psi^{(1)}\cdot\vec\psi^{(2)}-c]
=0
\label{psi1}
\end{equation}
and a similar equation for $\vec\psi^{(2)}(r)$, with
$\vec\psi^{(1)}(r)\leftrightarrow \vec\psi^{(2)}(r)$. Projecting these equation
to
$\hat a$ and to a perpendicular direction we can see that these projections
are consistent with (\ref{soln}). Thus, we have proven that ansatz
(\ref{higgsansatz}) combined with (\ref{soln}) and $\vec w =w\hat a$ are
consistent with the equations of motion. 

We are now in the position to be able to write down three scalar equations for
$w$, and the components of the Higgs fields. In
view of (\ref{soln}) we use only a total of two components for the two Higgs
fields. Note that a gauge rotation allows to rotate $\vec \psi_\bot$ into the
direction of one of the axes.  Therefore, we are able to deal with a single
perpendicular component only. Denoting the two independent components of the
Higgs fields by $\psi_{ a}$ and $\psi_\bot$ the three field equations are then 
\begin{equation}
\frac{3\,w'}{r}+w''-2e^2\Phi_0^2\left[w+\frac{1}{er^2}\right]
\psi_{\bot}^2
=0,
\label{weq2}
\end{equation}
\begin{equation}
\psi''_\bot+
\frac{1}{r}\psi'_\bot-e^2r^2\left(w+\frac{1}{e\,r^2}\right)^2
\psi_\bot-\frac{\Phi_0^2\,\psi_\bot}{2}\left[\lambda_1(\psi_\bot^2+\psi^2_{
a}-1) + \lambda_2(\pm\psi_{ a}^2\mp\psi_\bot^2- c)\right]
=0
\label{psiperp}
\end{equation} 
and
\begin{equation}
\psi''_{ a}+
\frac{1}{r}\psi'_{ a}-\frac{\Phi_0^2\,\psi_{ a}}{2}\left[\lambda_1
(\psi_a^2+\psi^2_\bot-1)
- \lambda_2(\pm\psi_a^2\mp\psi_\bot^2- c)\right]
=0.
\label{psia}
\end{equation}
 
 (\ref{weq2}), (\ref{psiperp}), and (\ref{psia}) do have nontrivial
solutions.  
  As an example, choosing the upper signs,\footnote{Numerical investigations
show that this choice leads to lower energies.} observe that the equations for
$\psi_\bot$ and $\psi_a$ decouple at  $\lambda_1=\lambda_2$.  (\ref{psia})
becomes independent of $\psi_\bot$ and of $w$ having a constant minimal energy
solution, $\psi_a=\sqrt{(1+c)/2}$.  Then (\ref{psiperp}) simplifies to
\begin{equation}
\psi''_\bot+
\frac{1}{r}\psi'_\bot-e^2r^2\left(w+\frac{1}{e\,r^2}\right)^2
\psi_\bot-\Phi_0^2\lambda\,\psi_\bot\left(\psi_\bot^2-\frac{1-c}{2}\right) 
=0.
\label{psiperp2}
\end{equation}
It is easy to recognize the system of equations (\ref{weq2})  and
(\ref{psiperp2}) as the rescaled version of the equations for the 2+1 dimensional
Abelian soliton (relativistic superconductor).~\cite{nielsen}~\cite{mandelstam}
If one sets $c=0$ then it is also equivalent to the equations found  in
Ref.~\cite{devega1}~\cite{devega2} Numerical investigations of those models show
 that these equations have a nontrivial solution satisfying the
appropriate boundary conditions. 

It is easy to visualize the motion of the vectors $\vec\psi_1$ and $\vec\psi_2$
as $r$ changes between $0<r<\infty$. At $r=0$ $\psi_{\bot}$ vanishes, so the two
vectors coincide.  They both point into the fixed direction of $\hat a$. Then, as
$r$ increases, $\vec\psi_1$ and $\vec\psi_2$ develop opposite components
perpendicular to $\hat a$, until these components reach the value
$\pm\sqrt{(1-c)/2}$ at $r=\infty$. At the same time, the vector potential,
being regular at the origin, behaves as $W_\mu= (1/e)\epsilon_{\mu i}x^i/r^2$ at
large values of $r$ corresponding to a finite magnetic flux along the $z$ axis,
$F=2\pi/e$. As expected, the vector potential is a pure gauge transformation at
infinity,
$\delta w_\mu=(-i/e)U^\dagger(\phi)\partial_\mu U(\phi).$

Solutions of the equation of motion at $\lambda_2\not=\lambda_1$ also exist. At
small $\lambda_1-\lambda_2$ one can calculate these solutions by perturbation
theory. In general, the boundary condition for $\psi_z$ at $r=0$ is that
$\psi_z(0)$=finite.  Our investigation of numerical solutions will be
presented in a future publication.~\cite{herat}

One more comment about our solution:  If we used different 
self-coupling for the two Higgs bosons then the solution $\vec\psi_{1\bot}=-\vec\psi_{2\bot}$ and $\vec\psi_1\cdot\hat a=
\vec\psi_2\cdot\hat a$ would not be admissible. Then one would get separate
equations for the four components of the two Higgs fields.  The dependence of
the Higgs fields on $r$ would become more complicated. 

In the remaining part of this letter we  would like to point out a
possible  relationship  between the vortices we discussed above with those
found using the method of center projection in lattice gauge
theories.~\cite{deldebbio}\cite{deldebbio2}\cite{alexandrou}\cite{second}\cite{faber}
Center projection is a method of realizing the Center Vortex Theory of
confinement~\cite{thooft0}~\cite{mack}~\cite{nielsen0} on lattices. Its aim is 
to extract the  degrees of freedom, most relevant for the nonperturbative
properties of nonabelian gauge theories, on a lattice.  Center projected
theories are theories of interacting vortices.  We shall point out that a 
certain class of center gauge fixing methods may be related to Higgs theories
with a pair of adjoint representation Higgs bosons.  This relationship offers a
possible way to define the so-called thick vortices,~\cite{deldebbio2} the
continuum analogue  of thin vortices (vortices with a cross section of a single
plaquette), appearing after center projection. 

Center Vortex Theory as dynamical model for confinement was proposed a long time
ago.~\cite{thooft0}~\cite{mack}~\cite{nielsen0}  This picture relies on the
condensation and percolation of magnetic vortices labeled by the elements of
the center of the gauge group, $Z_N$.
Vortices are one dimensional objects in three dimensional space and two
dimensional objects, like strings, in four dimensional spacetime.  They can be
contrasted to monopoles that are localized objects in space and one dimensional
objects, forming world lines, in four dimensional spacetime. Monopoles are are
the fundamental objects in the dual superconductor model of confinement by 't
Hooft.~\cite{thooft2} It is fairly easy to show that  a constant density of 
 percolated and randomly distributed magnetic vortex lines piercing Wilson loops
 on a lattice results in the area law for  Wilson loops and, consequently,
leads to confinement. 

The realization of these ideas on lattices has  been highly
successful in
$SU(2)$~\cite{deldebbio}~\cite{deldebbio2}~\cite{second}~\cite{alexandrou}, and
very recently, in $SU(3)$.\cite{faber} The first step of center projection
methods is
 to fix the gauge to retain only the symmetry corresponding to the center of the
gauge group.  The
gauge fixing is followed by projecting  the gauge fields to the center, $Z_N$,
leaving an interactive
$Z_N$ gauge theory.  

A variety of gauge fixing procedures has been used, with
the aim of transforming gauge fields to as close to the center of the group as
possible. Among others, the Maximal Center Gauge~\cite{deldebbio}~\cite{second}
(MCG) and the Laplacian Center Gauge~\cite{alexandrou} (LCG) are important to
mention. Both methods show convincingly that the resultant $Z_2$ (or $Z_3$)
gauge theory retains the essential nonperturbative properties of the original
nonabelian gauge theory, including confinement (with the correct coefficient in
the area law) and chiral symmetry breaking.  

  The  MCG~\label{deldebbio} method
maximizes the functional 
\begin{equation}
S_C=\frac{1}{4}\sum_{\mu,x} \left|{\rm Tr}U_\mu^V(x)\right|^2 
\label{mcg}
\end{equation}
over gauge transformations, $V(x)$.  Here
$U_\mu^V(x)=V(x)U_\mu(x)V^\dagger(x+\hat \mu)$ is the gauge transformed gauge
field on the lattice. 

It is easy to rewrite (\ref{mcg}) in terms of an adjoint
representation gauge transformation, which has the form\footnote{ Here we
use $SU(2)$ notations, though the generalization to $SU(N)$ is
straightforward.}
\[ V_{ij}(x)=\frac{1}{2}{\rm Tr}[V(x)\sigma_i V^\dagger(x)\sigma_j]
\]
and of the $SO(3)$ representation of the gauge fields
\[
U_{\mu ,ij}(x)=\frac{1}{2}{\rm Tr}[U_\mu(x)\sigma_i U_\mu^\dagger(x)\sigma_j]
\]
 as follows:
\begin{equation} S_V=\sum_{\mu,x,ijk} V_{ij}(x)U_{\mu,jk}(x)V_{ik}(x+\mu),
\label{mcg2}
\end{equation}
where the indices run from 1 to 3 in $SU(2)$.
Then one needs to  maximize (\ref{mcg2}) over all possible orthogonal
matrices
$V_{ij}(x)$.   

The rows (and columns) of the orthogonal matrix $V_{ij}(x)$ are
orthonormal. This constraint is relaxed and the 
largest eigenvalues and the corresponding eigenvectors, $v_i^{1,2}(x)$, of the
laplacian matrix
\[
\sum_\mu
[U_{\mu,ij}(x)\delta(x-y+\hat\mu)+U_{\mu,ji}(x-\hat\mu)
\delta(x-y-\hat\mu)-2\delta_{ij}\delta(x-y)]
\]
are found
in LCG.~\cite{alexandrou} After orthonormalizing  these vectors at every site
one can  find the gauge transformation generated by LCG. Note that it is not
necessary to find three orthogonal vectors, as the gauge is completely fixed by
 two columns of the matrix $V_{ij}(x)$.

Both of these gauge fixing procedures, 
when using only two columns of the adjoint representation gauge transformation,
 are equivalent to the minimization of the action of a gauge-Higgs model with
two adjoint representation Higgs bosons.  Introducing the notation
$V_{1i}(x)=\Phi^{(1)}_i(x)$ and $V_{2i}(x)=\Phi^{(2)}_i(x)$ the gauge fixing
term becomes the gauge invariant kinetic term of the two Higgs bosons.  Adding
self and mutual interaction terms one obtains the following Higgs action:
\begin{equation}
S_H[U,\Phi]=\sum_{x}\left\{\sum_{r,\mu,ij}\frac{1}{2}\Phi^{(r)}_i(x)U_{\mu,ij}(x)
\Phi^{(r)}_j(x+\mu)
+\frac{\lambda_1}{8}\sum_{r=1}^2[(\vec\Phi^{(r)})^2-1]^2+\frac{\lambda_2}{4}
[\vec\Phi^{(1)}\cdot\vec\Phi^{(2)}-c]^2\right\}.
\label{newhiggs}
\end{equation}
It should be understood that $S_H$ is used in a way to solve  $\partial
S_H/ \partial \Phi^{(r)}_i=0$ first, then to define the gauge transformation as
the one transforming the minimizing solution, $\Phi^{(r)}_i$, to a pre-determined
form, say
$\Phi^{(1)}_i+\Phi^{(2)}_i\sim
\delta_{i3}$ and  $\Phi_i^{(1)}\sim \alpha \delta_{i3}+\beta\delta_{i1}$ or to  
$\Phi_i^{(1)}\sim  \delta_{i3}$ and $\Phi_i^{(2)}\sim \alpha
\delta_{i3}+\beta\delta_{i1}$ (this latter prescription was followed by
Alexandrou et. al.~\cite{alexandrou})

It is easy to see that the Higgs gauge fixing term (\ref{newhiggs})
incorporates both MCG and LCG.  In the limit $\lambda_1\to \infty$,
$\lambda_2\to\infty$ and $c\to0$ the orthonormality of the two Higgs bosons is
enforced.  In the limit of $\lambda_1=\lambda_2=0$ all these constraints are
fully relaxed and the gauge fixing is just like in LCG. Our generalized gauge
fixing procedure is then defined by maximizing (\ref{newhiggs}) in the given
gauge field background and then choosing the gauge e.g. to rotate one of these
Higgs bosons parallel to the $z$ axis and the other one into the $xz$
plane.

The form of (\ref{newhiggs}) is tantalizing, as it offers a possible 
relationship between gauge theories with two adjoint representation Higgs
bosons and Center Gauge fixing.  Note that the condition $\partial
S_H/ \partial \Phi^{(r)}_i$ is identical to the one used in the Higgs theory to
find classical vortex solutions. As gauge fixing  does not affect
the physical gauge fields that were generated without gauge fixing,  the
extremum condition for the gauge field is not applicable. Still the vortices
appearing in the gauge fixed theory satisfy similar boundary conditions at
infinity and near the vortex core as in the Higgs theory.  Thus, one expects
that Center vortices and vortices in the Higgs theory are mathematically very
similar.  We conjecture that
the adjoint Higgs theory is a good laboratory for the analytic investigation of
center vortices, mostly for studying interaction and condensation of vortices. 

As vortices in center gauge fixed theories condense it would be of
considerable interest to investigate  multiple vortex configurations in Higgs
theories. Note however that e.g. in
$SU(2)$ the superposition of two vortices corresponds to the trivial homotopy
class and no stable double vortex solutions should exist.~\cite{thooft1} In
our view this does not mean that such configurations do not give
contributions to physical quantities.  Indeed, the generating function could be
dominated by condensates of vortices, partially due to phase space effects and
partially due to the difficulty of annihilating two `infinitely long' vortices
that are not parallel to each other. In a lattice gauge theory the `speed' of
creating these vortices can be in equilibrium with their `speed' of
annihilation. 

The author is indebted to Rohana Wijewardhana and Igor Shovkovy for
discussions and to Fidel Schaposnik for correspondence.  Partial support by the
U.S. Department of Energy through grant
\#DE FG02-84ER-40153 is gratefully acknowledged.

 \end{document}